\documentclass[9pt,twocolumn,twoside]{pnas-new}
\templatetype{pnasresearcharticle}
\usepackage{graphicx}
\usepackage{url}
\usepackage{bm,color,amsmath,amssymb}

\title{Using  posterior predictive distributions to analyse epidemic models: COVID-19 in Mexico City}

\author[a]{Rams\'es H. Mena}
\author[b]{Jorge X. Velasco-Hernandez} 
\author[c]{Natalia B. Mantilla-Beniers}
\author[c]{Gabriel A. Carranco-Sapi\'ens}
\author[d]{Luis Benet}
\author[e]{Denis Boyer}
\author[f,g,1]{Isaac P\'erez Castillo}

\affil[a]{Instituto de Investigaciones en Matemáticas Aplicadas y en Sistemas, Universidad Nacional Aut\'onoma de M\'exico, México
CDMX, Apartado Postal 20-726, 01000, México}
\affil[b]{Instituto de Matem\'aticas, Universidad Nacional Aut\'onoma de M\'exico Unidad Juriquilla 76230, Quer\'etaro, M\'exico}
\affil[c]{Facultad de Ciencias, Universidad Nacional Aut\'onoma de M\'exico 04510 CDMX,  M\'exico}
\affil[d]{Instituto de Ciencias F\'isicas, Universidad Nacional Aut\'onoma de M\'exico, Av. Universidad s/n, Col. Chamilpa,
C.P. 62210 Cuernavaca, Morelos, M\'exico}
\affil[e]{Departmento de Sistemas Complejos, Instituto de F\'isica, Universidad Nacional Aut\'onoma de M\'exico, Apartado Postal 20-364, 01000 CDMX, Mexico}
\affil[f]{Departmento de F\'isica Cu\'antica y Fot\'onica, Instituto de F\'isica, Universidad Nacional Aut\'onoma de M\'exico, Apartado Postal 20-364, 01000 CDMX, Mexico}
\affil[g]{London Mathematical Laboratory, 8 Margravine Gardens, London, W68RH, UK}

\leadauthor{P\'erez Castillo} 

\significancestatement{Epidemiological compartmental models are sloppy models: there exists a wide number of solutions in their parameter space that can fit equally well a given data describing the beginning of an epidemic. Under this condition, using only one point of this set of parameters to predict the evolution of the epidemic is clearly inappropriate. Using  posterior predictive distributions we can account for all possible epidemic curves compatible with the data and provide a more robust, and probabilistic forecast.}

\authorcontributions{All authors contributed to the discussions that led to this research, and corrections to the manuscript. The manuscript was mainly written by IPC and JXVH. The public database provided by the Mexican Government was processed by GACS and LB to extract the appropriate data. The calculations were performed by LB and IPC.}
\authordeclaration{Authors declare no competing interests.}
\correspondingauthor{\textsuperscript{1} isaacpc@fisica.unam.mx}


\begin{abstract}
Epidemiological models contain a set of parameters that must be adjusted based on available observations. Once a model has been calibrated, it can be used as a forecasting tool to make predictions and to evaluate contingency plans.
It is customary to employ only point estimators for such predictions.
However, some  models may fit the same data reasonably well for a broad range of parameter values, and this flexibility means that predictions stemming from such models will vary widely, depending on the particular parameter values employed within the range that give a good fit. 
When data are poor or incomplete, model uncertainty widens further. 
A way to circumvent this problem is to use Bayesian statistics to incorporate observations and use the full range of parameter estimates contained in the parameters' posterior distribution to adjust for uncertainties in model predictions.
Specifically, given the epidemiological model and a probability distribution for observations, we use the posterior distribution of model's parameters to generate all possible epidemiological curves, which are encapsulated in  posterior predictive distributions.
From these, one can extract the worst-case scenario and study the impact of implementing contingency plans according to this assessment. 
We apply this approach to the potential evolution of COVID-19 in Mexico City and assess whether contingency plans are being successful and whether the epidemiological curve has flattened.
\end{abstract}

\dates{This manuscript was compiled on \today}
\doi{\url{www.pnas.org/cgi/doi/10.1073/pnas.XXXXXXXXXX}}

\begin{document}

\maketitle
\thispagestyle{firststyle}
\ifthenelse{\boolean{shortarticle}}{\ifthenelse{\boolean{singlecolumn}}{\abscontentformatted}{\abscontent}}{}

\dropcap{D}ecember 2019 saw the start of an outbreak of pneumonia of unknown etiology in Wuhan, China.
This would be recognised as result of the disease provoked by a new coronavirus able to infect humans and transmit within human populations.
By January 23, Chinese authorities had taken severe measures to contain its spread: imposing travel bans, restricting mobility within Wuhan, isolating suspect and confirmed cases, banning gatherings and shutting schools and entertainment venues.
This did not prevent the virus from reaching several other countries and all regions of China quickly.
On January 30, with 7,711 confirmed cases in China and 83 in other countries, the World Health Organization declared SARS-CoV-2 a Public Health Emergency of International Concern.  \cite{Sitrep1,PHEIC}
	
Mexico confirmed its first cases of Covid-19 on February 27 in travellers returning from Italy to Sinaloa and Mexico City respectively.
On March 15, the Mexican National Committee for Safety in Health (Comit\'e Nacional para la Seguridad en Salud) announced the start, on March 23, of distancing measures to mitigate the transmission of COVID-19, declaring the start of the second phase of the epidemic.
Phase three would be declared nearly a month later, on April 21.
Distancing measures included suspension of all non essential activities of public, private and social sectors, and was initially planned to last until April 30, but was later extended until May 17 or May 30, depending on the local situation of every municipality of the country.
These measures were designed to lower disease incidence rates of COVID-19 and keep the number of hospitalized and critical cases manageable. \cite{CTD_inicio, Reporte_epi}

There are important reasons to expect that the number of actual infected cases in Mexico City, and the country at large, are larger than the reported ones.
The testing rate in Mexico is the lowest among the OECD countries \cite{COVID-2020} and the positivity rate for testing in Mexico City on the week ending on May 6, for example, ranges between 24.5\% to 41.7\% depending on the municipality \cite{CDMXdata}.
Both of these factors imply a likely large sub-reporting of cases.
The strain on the Health system is already important in Mexico City and other large population centers in the country.
To the day of submission of this work, the model used by the Federal Government's General Directorate of Epidemiology (Direcci\'on General de Epidemiolog\'ia) has not been publicly released.
There is no technical information available on the model's fundamental underlying (biological, statistical, mathematical) assumptions on contact rates, initial conditions, percentage of asymptomatic carriers, basic reproduction number among others.
As far as we are aware of, there is only one peer-reviewed published model on the Mexican case but this is centered on the analysis of the efficacy of the implementation of the mitigation strategies still in effect \cite{Acuna2020}.
There is, therefore the urgent need to count with alternative models able to project feasible scenarios of the epidemic in Mexico, in order to evaluate, compare and improve the expected trends, infection levels and public health strategies in view of the upcoming lifting (May 30) of the mitigation and social distancing measures in effect since March 23. 

This pandemic has shown that various parameter estimates vary wildly from country to country.
Thus, comparing fitted parameters between different countries to either disregard or confirm a particular model may be misleading.
It would be rather more sensible to run different models for a given population and compare the results for that particular setting.
The reasons as to why fitted parameters vary so much from country to country, one may speculate, may lay on the particular age distributions, risk factors, income, access to healthcare, social norms, climate, to mention but a few.

In this scenario, mathematical models are a natural tool for identifying what needs to be done in order to avoid saturating the healthcare system.
Models are commonly used to estimate, for example, the extent of the reduction in the effective transmission rate needed to control an epidemic.
However, most of these models fall in a category commonly known as \textit{sloppy models} \cite{Gutenkunst2007}.
These are models that depend on a large number of parameters and for which, once fitted to limited or noisy data, a broad range  of certain parameter values produce similarly acceptable fits.
This is clearly  disconcerting, since using different parameter estimates  one will surely  obtain widely different predictions from the same model, rendering its application to forecasting impractical, a problem which unfortunately is frequently overlooked at times when theoretical expectations and scientific rigour are directly needed.

Here, we present a Kermack-McKendrick type of model \cite{kermack1927} to evaluate the efficacy of the Sanitary Emergency declaration in containing disease spread in Mexico City, taking into account parameter uncertainty and data scarcity.
One way to tackle these uncertainties is to use a Bayesian approach and analyse whether the mitigation measures are being effective and what are the worst-case scenarios to be expected.
Specifically, we introduce the predictive posterior distribution for epidemiological models. This allows us to analyse a full spectrum of scenarios, thus enabling us to determine whether the response is being appropriate in order to avoid the collapse of the healthcare system.
Instead of accurately calibrating models with data, which is a difficult task to carry out from the short time series of the early stage of an epidemic, we focus on the effects of parameter variability in the model's predictions.

%%%%%%%%%%%%%%%%%%%%%%%
\section{On epidemiological models}
\label{sec:model}
%%%%%%%%%%%%%%%%%%%%%%%
The basic idea of epidemiological compartmental models is to split the host population (often assumed to be of constant size $N$) into $r$ compartments corresponding to states of the infection, so that $\mathcal{N}_a(t)$ indicates the population in state $a=1,\ldots,r$. 
We thus introduce vector $\bm{\mathcal{N}}(t)=(\mathcal{N}_1(t),\ldots,\mathcal{N}_r(t))$ and assume the epidemic to follow a set of nonlinear ODEs
\begin{equation}
\frac{d\bm{\mathcal{N}}(t)}{dt}=\bm{\mathcal{F}}[\bm{\mathcal{N}}(t),\bm{\theta}]\,,
\label{eq:1}
\end{equation}
where $\bm{\theta}=(\theta_1,\ldots,\theta_{p})$ is a set of $p$ parameters of the model.
Let $\bm{\mathcal{N}}(t,\bm{\theta})$ denote the solution for the set of equations \eqref{eq:1} given the parameters $\bm{\theta}$. Examples of simple epidemiological models are the Susceptible-Infected-Recovered (SIR) or the Susceptible-Exposed-Infected-Recovered (SEIR) models, for which the states are $\bm{\mathcal{N}}=(S,I,R)$ or $\bm{\mathcal{N}}=(S,E,I,R)$, respectively.
More realistic models, as the one we will use here, with the aim  to estimate disease toll and burden, introduce additional states to follow hospitalized and critically-ill patients.

Suppose now  that we have an observational dataset $\mathcal{D}\equiv\{\bm{\mathcal{N}}^{({\rm obs})}(t)\}_{t=0}^{t_{\max}}$, possibly with an observational time- and compartmental-correlation matrix.
From here we can derive the likelihood $P(\mathcal{D}|\bm{\theta})$ of observing this dataset given a set of parameters.
Using Bayes' rule, the posterior distribution of the parameters given the dataset is simply $P(\bm{\theta}|\mathcal{D})\propto P(\mathcal{D}|\bm{\theta})P_{0}(\bm{\theta})$ where $P_0(\bm{\theta})$ is the  prior distribution of the parameters.
The standard way to calibrate the model is to find the set of parameters, denoted here as $\bm{\theta}^\star$,  which maximizes the posterior distribution $P(\bm{\theta}|\mathcal{D})$, that is, $\bm{\theta}^\star=\text{arg max}_{\bm{\theta}} P(\bm{\theta}|\mathcal{D})$.
These are sometimes referred to as \textit{maximum a posteriori} (MAP) estimators.
When the  prior distribution is flat, and the posterior distribution exists, $\bm{\theta}^\star$ coincides with the maximum likelihood estimator.
Once the model has been calibrated using this point estimator, the evolution of the epidemic is given by $\bm{\mathcal{N}}(t,\bm{\theta}^\star)$, which can then be used to make predictions. 

Unfortunately, this method tends to fail for the so-called \textit{sloppy models} \cite{Gutenkunst2007}, because the variances in parameter calibration can be rather large in certain directions of the parameter space, particularly when using data only from the beginning of the epidemic curve.
As a result, there is large uncertainty in the conditions leading to the desired state, which renders this deterministic approach inadequate as a forecasting tool to e.g. implement contingency plans.
A full Bayesian approach considers the uncertainty captured by the whole posterior distribution $P(\bm{\theta}|\mathcal{D})$, and not only the deterministic point estimator $\bm{\theta}^\star$.
From this principle we can introduce various \textit{posterior predictive distributions}.
We start by considering the posterior predictive compartmental distribution given by:
\begin{equation}
    P(\bm{n},t|\mathcal{D})=\int d\bm{\theta}P(\bm{\theta}|\mathcal{D})P[\bm{\mathcal{N}}(t,\bm{\theta})=\bm{n}\mid \bm{\theta}]\,,
    \label{eq:2}
\end{equation}
where  $P[\bm{\mathcal{N}}(t,\bm{\theta})=\bm{n}\mid \bm{\theta}]=\delta[\bm{n}-\bm{\mathcal{N}}(t,\bm{\theta})]$, since the evolution equations \eqref{eq:1} modelling the epidemic are deterministic.
Here $P(\bm{n},t|\mathcal{D})=\text{Prob}(\bm{\mathcal{N}}(t,\bm{\theta})=\bm{n}\mid \mathcal{D})$ corresponds to the probability of observing  a given value of state $\bm{n}=({n}_1,\ldots, n_r)$ at time $t$ given the data set $\mathcal{D}$.
Clearly, if $P(\bm{\theta}|\mathcal{D})$ has a marked peak around $\bm{\theta}^\star$, with the extreme case being $P(\bm{\theta}|\mathcal{D})=\delta(\bm{\theta}-\bm{\theta}^\star)$, then $P(\bm{n},t|\mathcal{D})$ evolves deterministically according to $\bm{\mathcal{N}}(t,\bm{\theta}^\star)$, that is $ P(\bm{n},t|\mathcal{D})=\delta\left[\bm{n}-\bm{\mathcal{N}}(t,\bm{\theta}^\star)\right]$, which then recovers the previously mentioned standard approach.
However, if the posterior distribution $P(\bm{\theta}|\mathcal{D})$ is spread wide, so will be $P(\bm{n},t|\mathcal{D})$.
Thus, we need to consider the whole distribution $P(\bm{n},t|\mathcal{D})$ as a forecasting tool, and use it to analyse the implementation and impact of contingency plans.

Generally, we expect the posterior predictive distribution $P(\bm{n},t|\mathcal{D})$ to have a compact support, since the host population is taken to be constant.
With this in mind, we will denote as $\bm{\Omega}^{({\rm low})}(t)$ and $\bm{\Omega}^{({\rm up})}(t)$ its lower and upper boundaries, respectively, that is,  $P(\bm{n},t|\mathcal{D})$ is zero for $\bm{n}\not\in[\bm{\Omega}^{({\rm low})}(t),\bm{\Omega}^{({\rm up})}(t)]$. 
The two boundaries, ${\Omega}_a^{({\rm low})}(t)$ and ${\Omega}_a^{({\rm up})}(t)$, which correspond fairly intuitively to the lower and upper envelopes of all possible epidemiological curves ${\mathcal{N}}_a(t,\bm{\theta})$ with $\bm{\theta}$ drawn from $P(\bm{\theta}|\mathcal{D})$, can be understood in epidemiological terms as the best- and worst-case scenarios of the epidemic for state $a$ at time $t$, respectively.
Thus, they are fairly useful to determine the impact on a healthcare system.
For instance, if we were to have a compartment $C$ modelling critically-ill patients, the corresponding upper boundary $\Omega_C^{({\rm up})}(t)$ gives a bound for the worst-case scenario.
Thus, if a particular healthcare system has a given maximum capacity, denoted here as $\mathcal{B}$ (e.g. total Intensive Care Units available) to treat critically-ill patients, then having $\Omega_C^{({\rm up})}(t)>\mathcal{B}$ at some point indicates that the healthcare system has demands exceeding its capacity.
A careful, and  successful, contingency plan must consider the worst possible outcome of the epidemic, so that implemented measures guarantee that $\Omega_C^{({\rm up})}(t)<\mathcal{B}$.

Equally important is to derive the posterior predictive distribution of times at which the epidemic curve will peak.
Indeed, let $t^{(a)}_{\rm peak}=\text{argmax}_{t} \mathcal{N}_a(t,\bm{\theta})$ be the time at which the epidemic  reaches its peak for compartment $a$, and let us further denote $\bm{t}_{\rm peak}=(t^{(1}_{\rm peak},\ldots,t^{(r)}_{\rm peak})$.
The corresponding posterior predictive distribution of times at which the peaks occur reads:
\begin{equation}
    P\left(\bm{t}_{\rm peak}\big|\mathcal{D}\right)=\int d\bm{\theta}P(\bm{\theta}|\mathcal{D})\delta\left[\bm{t}_{\rm peak}-\text{argmax}_{t}\, \bm{\mathcal{N}}(t,\bm{\theta})\right]\,.
   \label{eq:3}
\end{equation}

Notice that one would be tempted to predict the peak of the epidemic based on  [\ref{eq:2}] by first obtaining the mean value for a given compartment, $\langle n_a(t)\rangle_{P(\bm{n},t|\mathcal{D})}$, and then look for the time at which the mean curve peaks, $\text{arg max}_t=\langle n_a(t)\rangle_{P(\bm{n},t|\mathcal{D})}$.
Clearly this is not necessarily equal to $\langle t^{(a)}_{\rm peak}\rangle_{ P(\bm{t}_{\rm peak}|\mathcal{D})}$, so it is more appropriate to use the  posterior predictive distribution of times, to correctly assess the probability for the peak to occur at a given time.

%%%%%%%%%%%%%%%%%%%%%%%%%%%%%%%%%%%%%%%%%%
\section{Model selection, and resulting analysis for COVID-19 in Mexico City}
\label{sec:calibration and results}
%%%%%%%%%%%%%%%%%%%%%%%%%%%%%%%%%%%%%%%%%
%%%%%%%%%%%%%%%%%%%%%%%%%%%%%%%%%%%%%%%%%%
\subsection{Model selection}
%%%%%%%%%%%%%%%%%%%%%%%%%%%%%%%%%%%%%%%%%
For the compartment  model used to analyse the data of COVID-19 for Mexico City, we have chosen to follow the one used in \cite{Maslov2020,AksamentovNeher2020} (and references therein).
Here, susceptible individuals $S$ become exposed ($E$) to the  virus through contact with infected individuals $I$.
Exposed individuals progress towards the symptomatic state $I$ within an average time  $\tau_\ell$.
As usual, mixing is assumed to be homogeneous.
Infected individuals $I$ cause an average of $R_0$ secondary infections over their infectious period.
After an average time $\tau_i$ (days), infected individuals either recover or progress towards hospitalization.
In turn, hospitalized individuals $H$ either recover or worsen towards a critical state after a time $\tau_h$.
Critical individuals $C$ allow us to  model ICU demand. 
They either return to state $H$, or die, moving to $D$, after a time scale $\tau_c$.
Recovered individuals $R$ are  assumed to be immune.
The dynamics of this model is given by the following set of differential equations:
\begin{eqnarray}
\frac{dS(t)}{dt}&=&-\beta(t) \frac{S(t)I(t)}{N}\label{eq:s}\\
\label{Sdot}
\frac{dE(t)}{dt}&=&\beta(t)\frac{S(t)I(t)}{N}-\frac{E(t)}{\tau_\ell}\\
\label{Edot}
\frac{d I(t)}{dt}&=&\frac{E(t)}{\tau_\ell}-\frac{I(t)}{\tau_i}\\
\label{Idot}
\frac{dH(t)}{dt}&=&(1-m)\frac{I(t)}{\tau_i}+(1-f)\frac{C(t)}{\tau_c}-\frac{H(t)}{\tau_h}\\
\label{Hdot}
\frac{d C(t)}{dt}&=&c \frac{H(t)}{\tau_h}-\frac{C(t)}{\tau_c}\\
\label{Cdot}
\frac{dR(t)}{dt}&=&m \frac{I(t)}{\tau_i}+(1-c)\frac{H(t)}{\tau_h}\\
\label{Rdot}
\frac{dD(t)}{dt}&=&f \frac{C(t)}{\tau_c}\,.\label{eq:d}
\label{Ddot}
\end{eqnarray}
The fraction of infections that are mild is $m$, the fraction of cases that turn critical is $c$, and the fraction of critical cases with fatal  outcome is $f$.
Other variants of the model consider, for instance, a recovery time for mild infections which is different from $\tau_i$, or a  fraction of those infected that are asymptomatic.
Equations [\ref{eq:s}-\ref{eq:d}] provide a relatively simple description of epidemic dynamics, including entry to and exit from the hospital, that allows us to focus on the number of hospitalized and critical cases, and foresee whether health services will be saturated.
The transmission parameter in the model is taken to be
\begin{equation}
\beta(t)=\frac{R_0M(t)}{\tau_i}
\label{beta_t}
\end{equation}
where $R_0$ is the basic reproduction number, and $M(t)$ captures the mitigation measures. 
While generally speaking pathogens affect populations in an uneven way, due to heterogeneity in the risk experienced by age, comorbidities or other factors (e.g. behaviour, nutrition and so on), for simplicity we assume a population homogeneous in all respects. 
A generalization to include how a particular age distribution affects model evolution is straightforward \cite{Maslov2020}, and is ongoing work.

%%%%%%%%%%%%%%%%%%%%%%%%%%%%%%%%%%%%%%%%%%
\subsection{Analysis and results for COVID-19 in Mexico City}
%%%%%%%%%%%%%%%%%%%%%%%%%%%%%%%%%%%%%%%%%
We applied this approach to study the evolution of the spread of SARS-CoV-2 in Mexico City using the public database provided by the Federal Health Secretariat \textit{Secretar\'ia de Salud Federal} corresponding to May 7 \cite{DatosAbiertos,DatosAbiertosGithub}.
From here, we have considered the data starting on February 27 (which we denote as $t=0$) up to April 29, to consider for delays in the reporting of cases due to delays in requests for medical attention, reporting or test confirmation.
The database allows extraction of incidence time series (new cases), as well as those newly hospitalised and critically-ill (complicated hospitalisations including the use of mechanical ventilators).
It also includes the total number of deceased patients, patients that were lab confirmed of being infected of SARS-CoV-2, as well as suspect cases awaiting results of RT-PCR tests.
Note that new cases do not correspond to the number of cases in each compartment, a piece of information which is not in the data.
When calibrating the model we have considered a cautious approach to add half of the suspect cases to those confirmed for each of the aforementioned compartment, based off the estimations of the positivity test rate for suspect cases.
Clearly, not all suspect cases will be confirmed as SARS-CoV-2, since this epidemic is concurrently happening with other seasonal diseases and therefore we are somewhat describing an aggregate of all the seasonal epidemics currently going on in Mexico City, with more weight towards SARS-CoV-2. 
However, we believe it is important to include some of those suspected cases since they may add to the demand on the healthcare system.
Considering that the contingency plan was first activated March 23, we assume the mitigating function $M(t)$ to equal one before that date, and a constant $0\le\gamma\le 1$ (considered as a parameter), after that day.

All in all, given the data, we need to determine the parameters' posterior distribution $P(\bm{\theta}|\mathcal{D})$ (see Supporting Information for details on the model's calibration).
Notice that while the data gives some of the initial conditions for some compartments, we do not have information for others, in particular, for the initial conditions  $S(0)$ and $E(0)$. Thus we consider these to be also parameters of the model.

\begin{figure*}[ht]
    \centering
    \includegraphics[width=5.8cm,height=4cm]{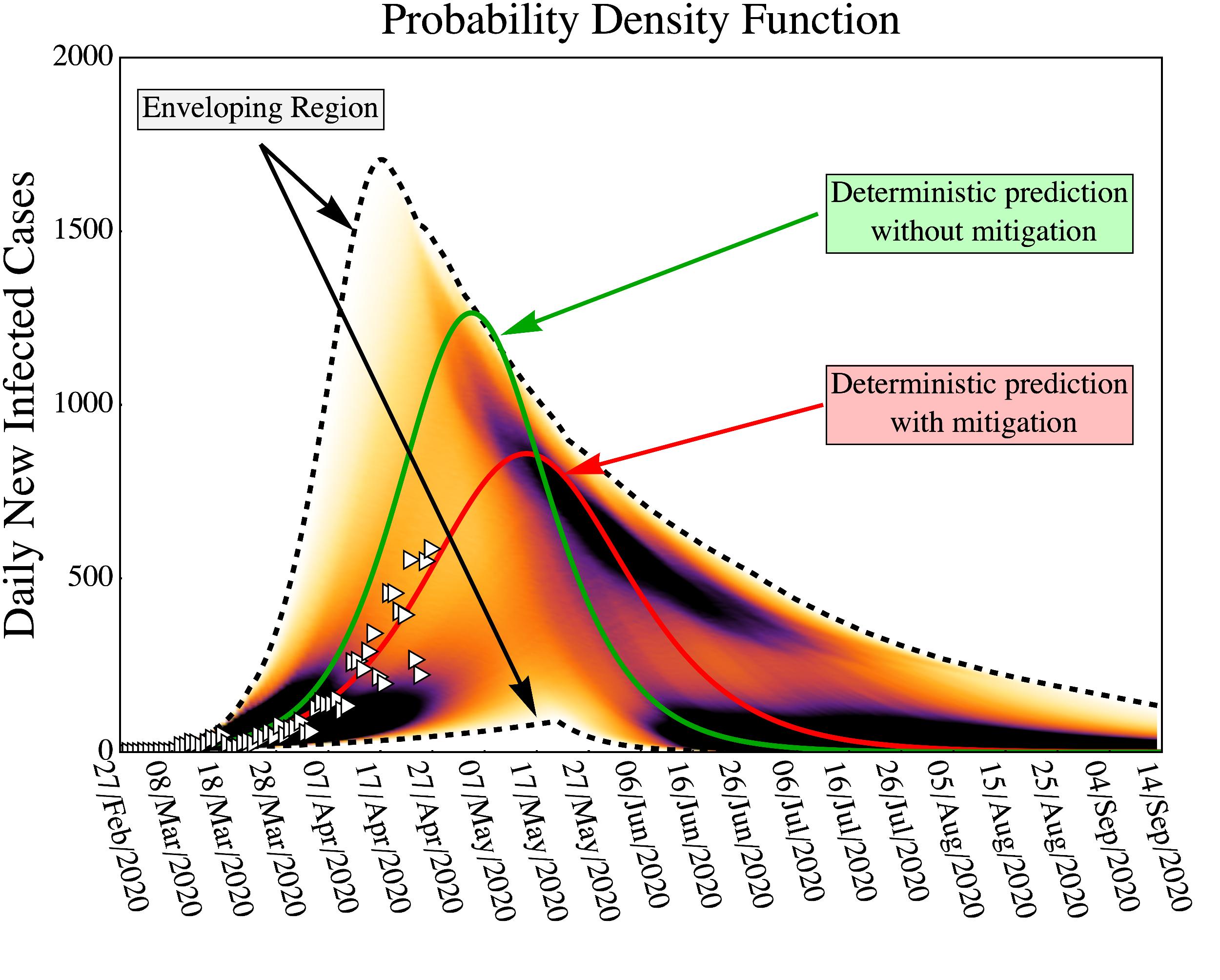}\includegraphics[width=5.8cm,height=4cm]{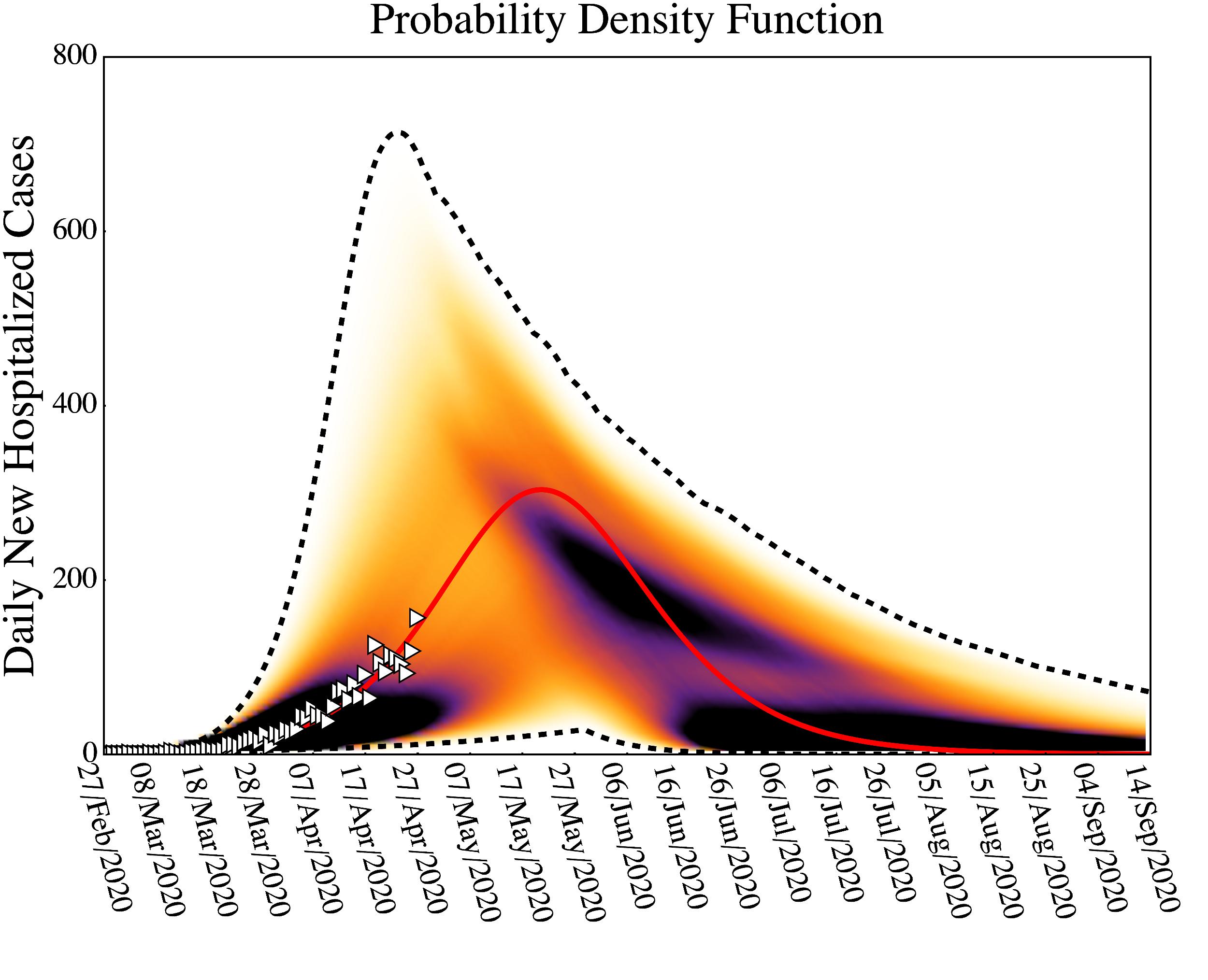}
    \includegraphics[width=5.8cm,height=4cm]{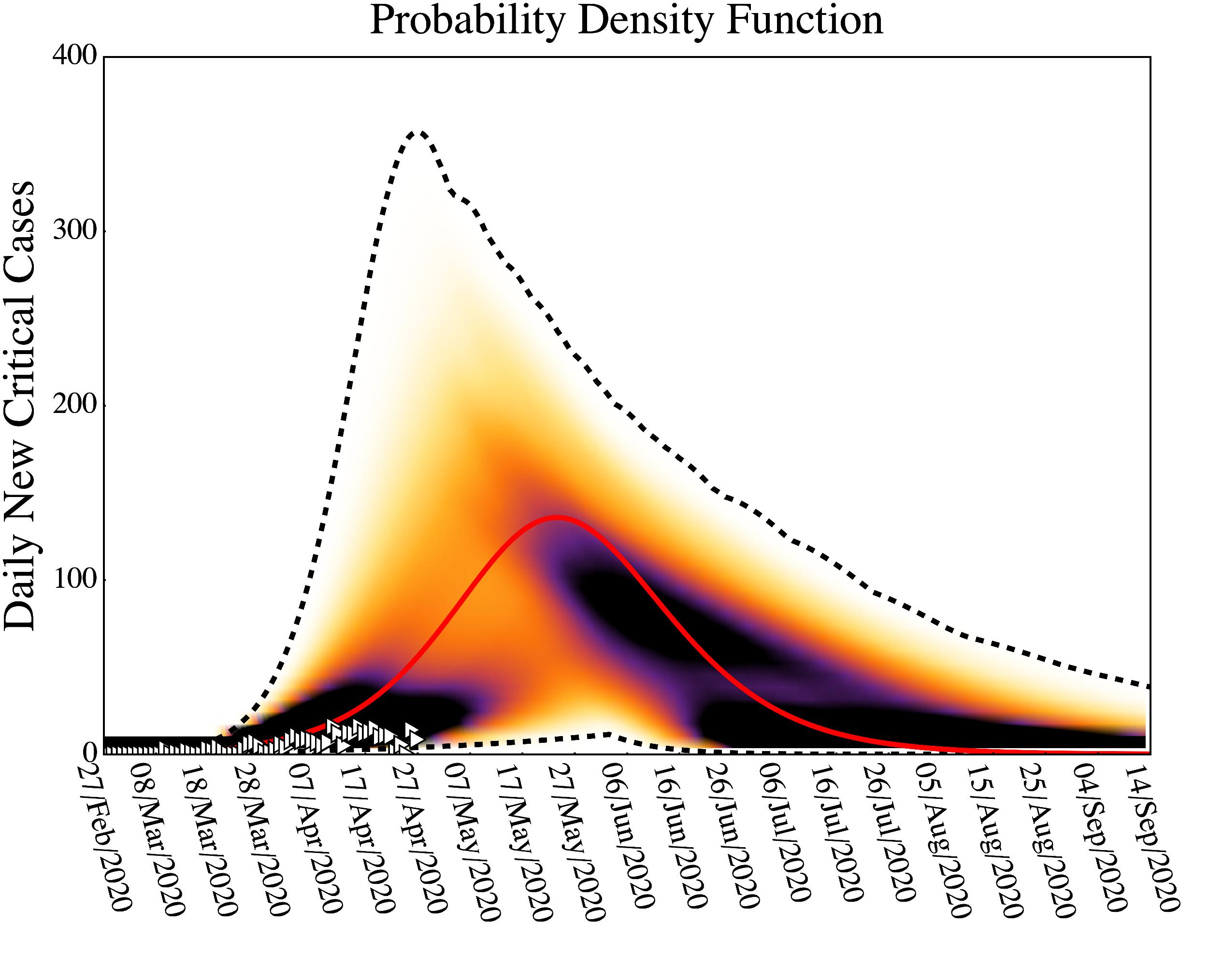}\\
    \includegraphics[width=5.8cm,height=4cm]{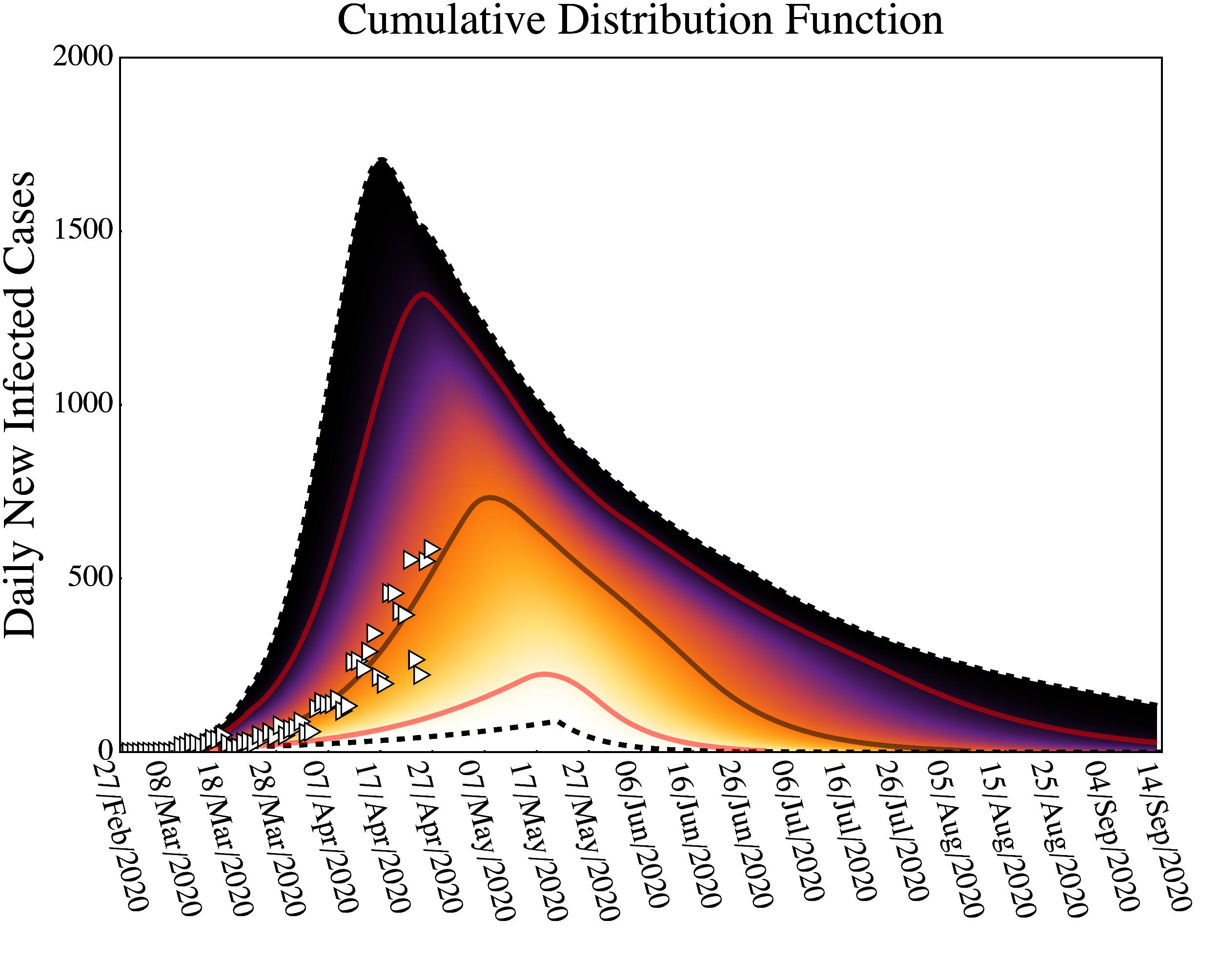}
    \includegraphics[width=5.8cm,height=4cm]{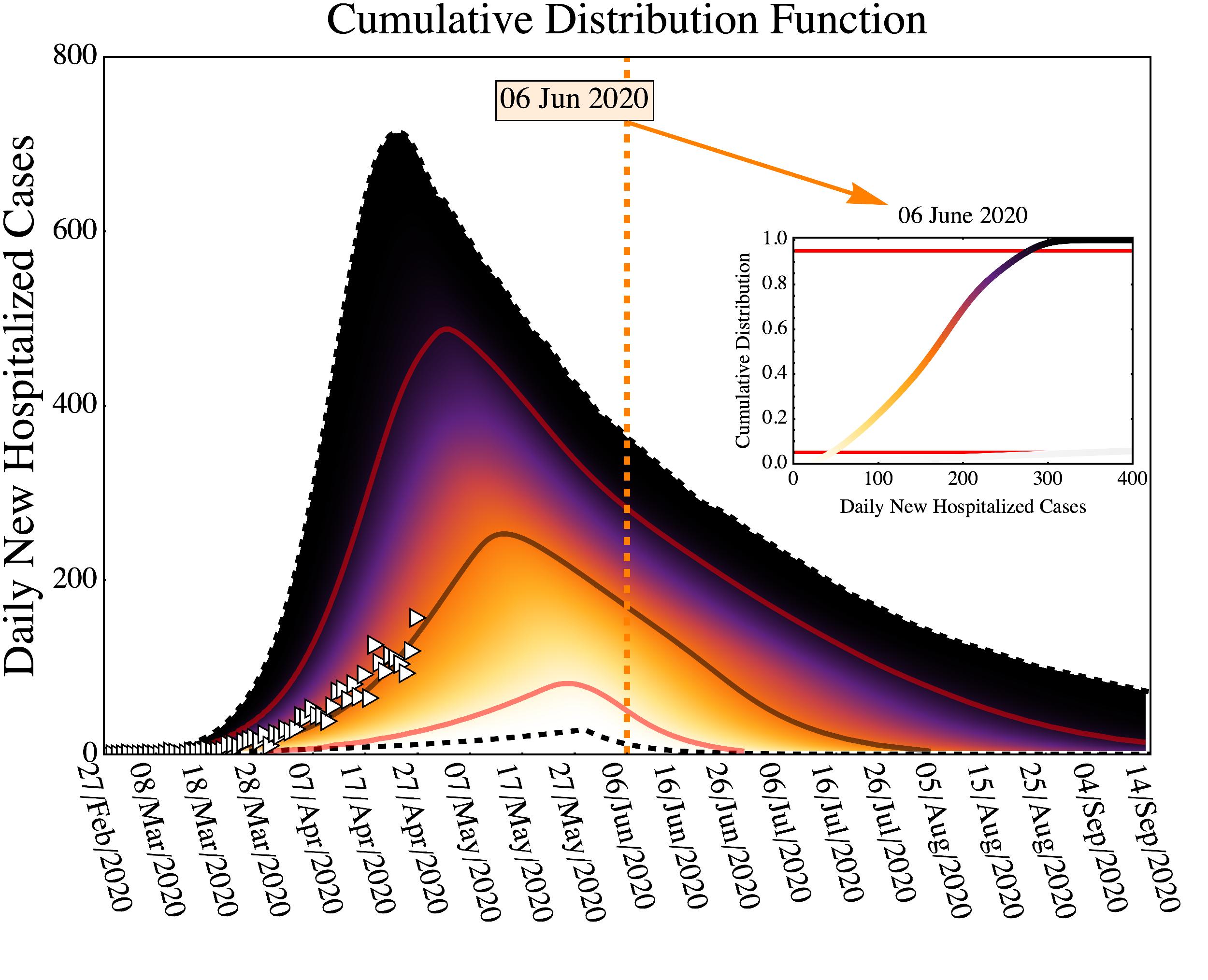}
    \includegraphics[width=5.8cm,height=4cm]{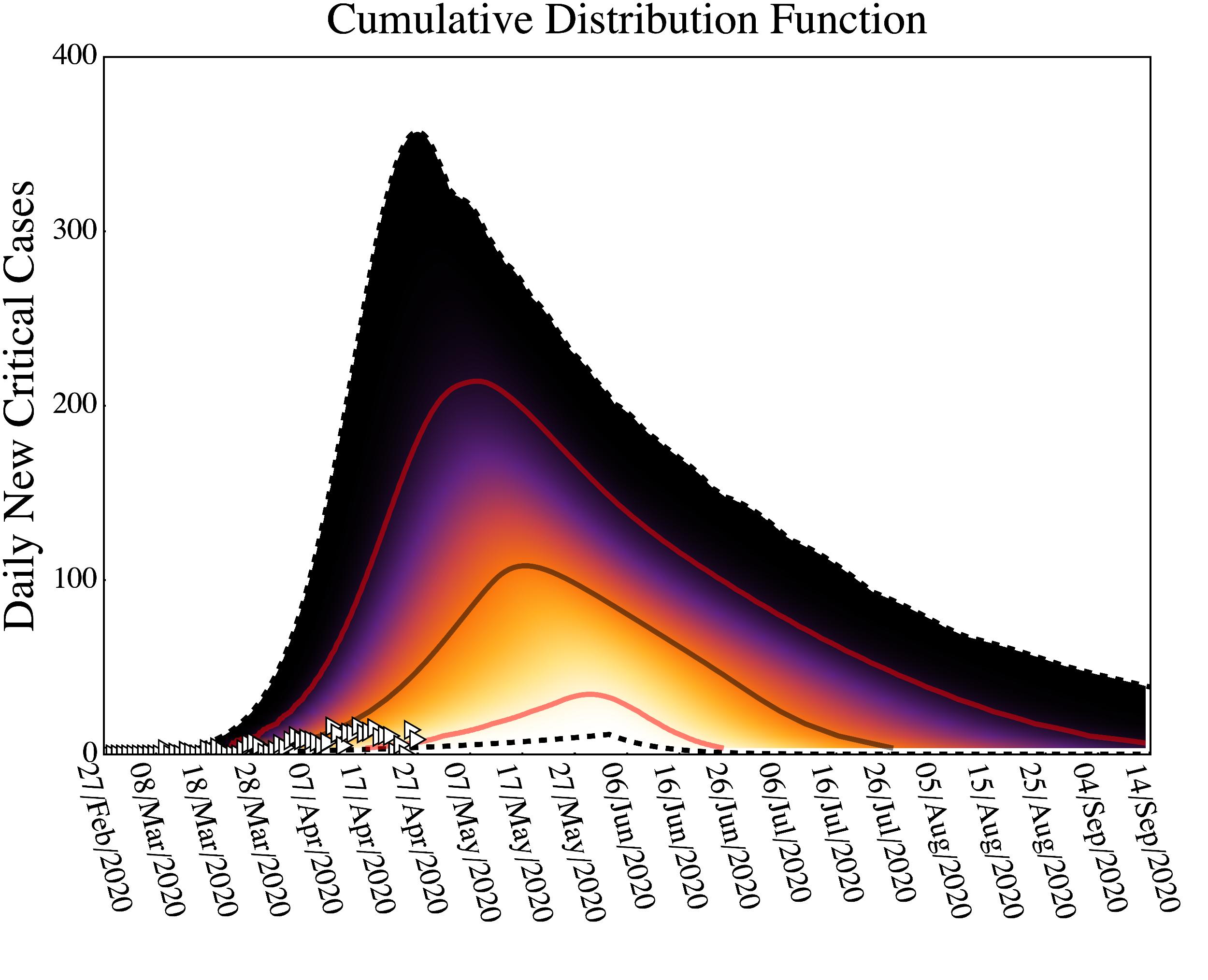}\\
      \includegraphics[width=5.8cm,height=4cm]{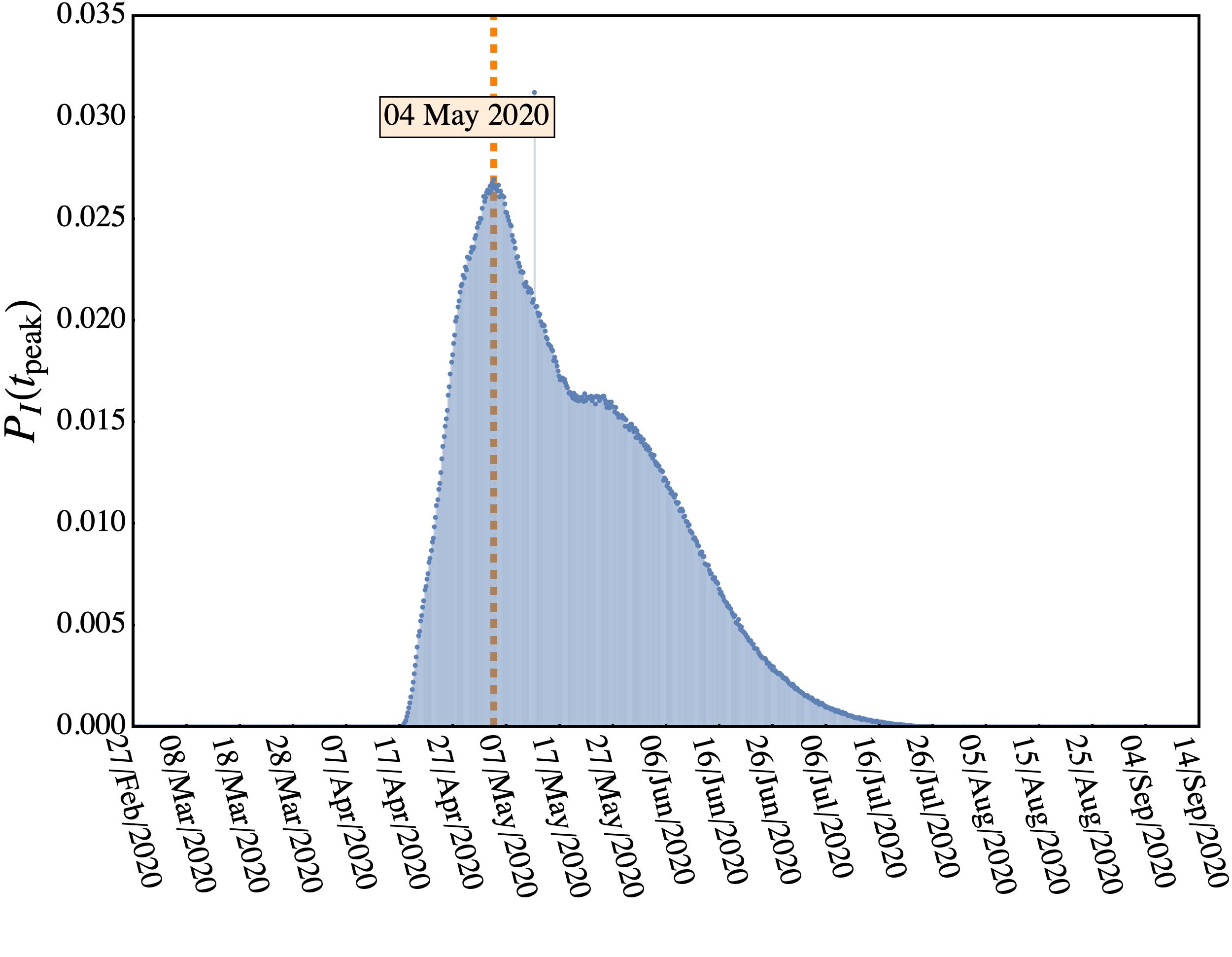}
    \includegraphics[width=5.8cm,height=4cm]{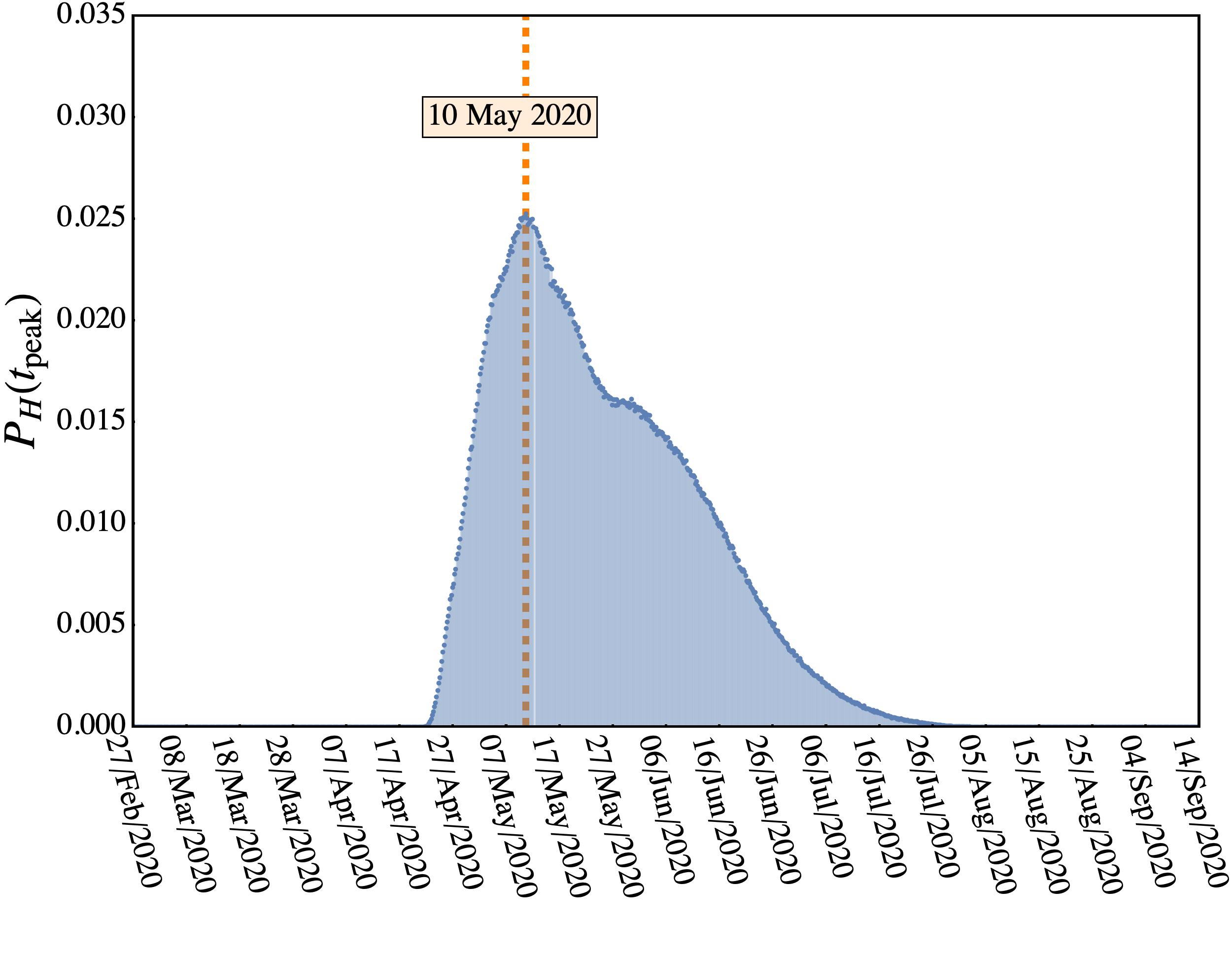}
    \includegraphics[width=5.8cm,height=4cm]{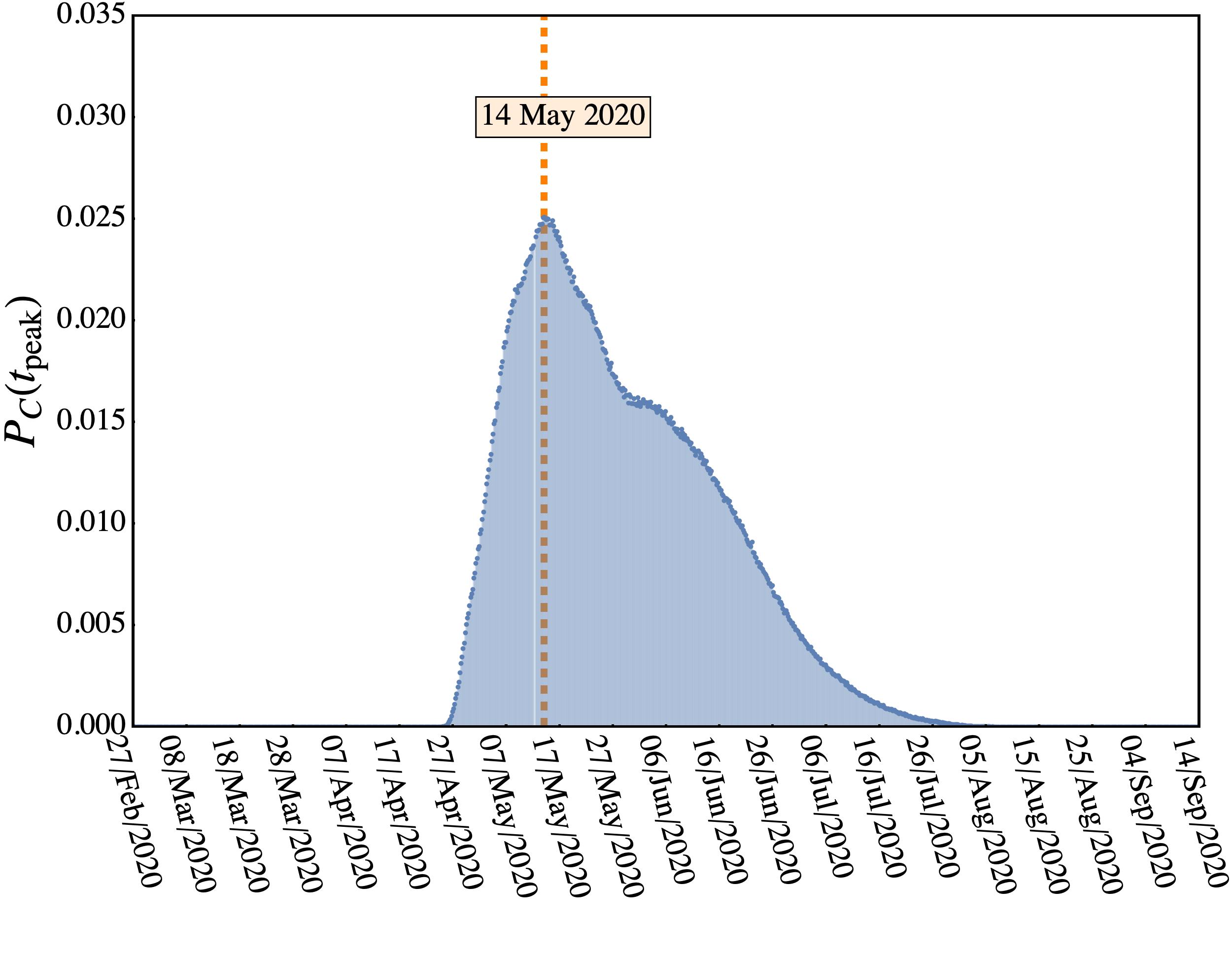}
    \caption{
        Top row: results for newly infected, hospitalized, and critical cases per day.
        In the first plot we indicate the meaning of each curve: the solid red line corresponds to the deterministic prediction with the mitigation plan; the solid green line is the corresponding deterministic prediction without mitigation.
        Dashed, solid lines comprise the envelope for the predictive posterior, while density plots give the actual value of the predictive posterior for the corresponding compartment in each plot; darker regions correspond to the accumulation of the  epidemic curves.
        Finally, white markers correspond to data for Mexico City. Middle row: in this case the density plots correspond to the CDF for newly infected, hospitalized, and critical cases per day.
        The solid black lines correspond to the median curve, while the lower and upper solid red lines are the  5\% and 95\% percentiles, respectively.
        In the middle figure in this row, we have added an inset plot, showing a cut of the CDF for a particular day.
        Bottom row: Posterior predictive distribution of times at which the epidemic curves will peak for daily new infected (left panel), hospitalized (middle panel), and critical (right panel) cases.
        The vertical lines indicate the data at which the peak would have occurred without mitigation.
        Note that the offset in dates among the three compartments in these plots can be roughly understood as the mean time a patient takes to become hospitalized from becoming infected and to become critically ill from being hospitalized.
        }
    \label{fig:1}
\end{figure*}

Once we have estimated $P(\bm{\theta}|\mathcal{D})$, we can use the expressions [\ref{eq:2}] and [\ref{eq:3}] to estimate the predictive posteriors.
All the results from these distributions are summarised in the plots appearing in Fig. \ref{fig:1}.

The first row of plots in this figure shows the resulting predictive posterior for daily new cases of infected, hospitalized and critically-ill patients.
In all cases, the solid red line correspond to the calibrated model with $\bm{\theta}^\star$, the white markers correspond to observational data, the density plots are the values of the predictive posterior distribution and, finally, the lower and upper dashed black lines delimit the enveloping region of all possible predictive scenarios for each incidence. 
%are the best and worst scenarios for each incidence. 
This set of plots are fairly informative and it is worth discussing them in detail.
We first notice that by using the parameter's posterior distribution, the deterministic solution (indicated by a solid red line for each frame in the first row), the solutions spread fairly widely, with all possible epidemic curves encapsulated by the dashed black lines.
Thus, the deterministic solution is very sensitive to parameter changes, which makes it unsuitable as a forecasting tool by itself.
Secondly, the density plots show that certain epidemic curves tend to accumulate in specific (dark) regions.
Interestingly enough, there is an increment in the density of curves symmetrically distributed above and below the deterministic curve at the beginning of the epidemic.
It turns out that the increased density above the red solid line corresponds to the epidemic that would have resulted if no contingency plan had been implemented.
The latter is indicated by a solid dark green line only on the first plot. We thus conclude that the contingency plan was successful, albeit mildly, managing to flatten the curve and shift its peak to the right.
Actually from the parameters' posterior distribution, one can show that prior the activation of the contingency plan, the basic reproduction rate $R_0$ was $2.48$, which was lowered to the value $2.03$ once the plan was activated on March 23.
Similarly, the increased density of curves below the deterministic line indicate what would have happened if the mitigation had been more successful.
We finally observe that the deterministic curve for daily new critically-ill patients obtained is above the data,  suggesting that we are overestimating the total toll for the number of deceased patients predicted by our analysis. 

The colors of the middle row of Fig. \ref{fig:1} indicate the cumulative distribution function for newly infected, hospitalized, and critical cases per day.
Thus, in this case, the color scale in the density plot goes from zero (white) to one (black).
The solid black line on these three plots corresponds to the median curve, while the lower and upper curves (marked in solid red lines) are the 5\% and 95\% percentiles.
In other words, the probability that all epidemiological curves generated by the calibrated model are comprised between the two solid red lines is 90\%.
Notice that one shortcoming of using only point estimators in compartmental models is that they yield epidemic curves which are fairly symmetric around their maximum, a feature that is not observed in the data from other countries, where fattening of the tails after the maximum is instead discerned.
However, by using Bayesian statistics one can produce more realistic epidemic curves, with fattened tails, as can be appreciated in the median curves reported in the second row of Fig. \ref{fig:1}.

We can similarly explore the posterior predictive distribution of times at which the peak of the epidemic occurs.
These are shown at bottom row of Fig.~\ref{fig:1} for newly infected, hospitalized and critically-ill patients per day, which were obtained according to Eq. [\ref{eq:3}].
These distributions are again very informative: in all of them the peak corresponds to the day at which the epidemic curve would have peaked with no contingency plan.
Interestingly enough, the support of the distribution of times is compact, meaning  that one could  provide a rather hard and robust interval within which peak actually happens, admittedly rather large.
We can also provide the mean date for the peak to occur.
For instance the mean date for new infected cases is May 18, with a standard deviation of 17 days. 
One may argue that having a rather large standard deviation does not provide informative predictions for the peak of the epidemic.
However, notice that the total span in days of the evolution of the first wave of the epidemic, until it finishes, is around 9 months.
A similar analysis follows for the other two posterior distributions for daily new hospitalised and critical cases.

%%%%%%%%%%%%%%%%%%%%%%%%%%%%%%%%%%%%%%%%%%
\section{Conclusions and future work}
\label{sec:conclusions}
%%%%%%%%%%%%%%%%%%%%%%%%%%%%%%%%%%%%%%%%%
Contingency plans based on epidemiological models must be analysed and carried out very carefully.
Even with fairly accurate observational data, the importance of stochasticity inherent to the start of an epidemic means that parameter estimates based on data from the beginning of an outbreak will be quite uncertain.
In turn, models parametrised with such data will carry great uncertainty in longer term forecasts.
On the other hand, this uncertainty can be quantified using techniques from Bayesian statistics, which may then be used to consider worst-case scenarios.

Although the model analysed here is simple, the main conclusion of this work is that extrapolating results without accounting for sensitivity to changes in parameters can result in predictions way off the mark.
We believe that the same conclusion would hold for more detailed models, e.g., those which include specific details of the population, since most of them are also sloppy. 

With regards to the mitigation measures implemented in Mexico City, our results show that they have so far managed to flatten the curve  moderately, thus shifting the peak for newly infected cases per day to the right, to a date around June 1.
However, this and other compartmental models, are rather sensitive to parameter calibration.
Access to richer data containing more epidemiological and clinical information would help to better control model predictions. 

Control of the epidemic curve of SARS-CoV-2 in Mexico City requires evaluating the mitigation strategies that are, to date, being implemented in the country.
Mathematical models are central to this effort, but certain conditions need to be considered and evaluated for their efficient application.
Mexico has the lowest testing rate among the OECD countries \cite{COVID-2020}.
A high testing rate is recommended to adequately plan when to lift mitigation measures now in place.
Moreover, testing is necessary to estimate the true size of the epidemic.
In Mexico, several hundreds of Health Units constitute the country's sentinel surveillance system where cases are detected and followed to identify possible contacts of that case and other relevant information \cite{SSVigil, SSVigil_nuevo}.
A case detected by symptomatic surveillance has to be confirmed by testing, but, to obtain a concrete, workable estimation of the epidemic, tests must be widely applied to the general population, not only to suspect cases already detected by the surveillance system.  

The positivity test rate for SARS-CoV-2 in the various municipalities of Mexico City was around 20\%-40\% on May 8, 2020 \cite{CDMXdata}.
This high positivity rate and the limited number of tests currently performed may prevent obtaining an accurate estimate of both the epidemic size and the true growth rate of the epidemic including the determination of the days where the epidemic peak is occurring; in particular, identification of the time of maximum incidence may be confounded.
Since tests are insufficient and, for the particular situation of the Mexican economy, increasing the testing rate is unfeasible, mathematical modeling projections can help to evaluate different scenarios that are consistent with the observed trend of the epidemic curve.
Our model provides projections based on confirmed cases corrected for under-reporting that put the more likely dates of maximum incidence towards the end of May or early June, 2020.
Earlier dates are possible, too, but with lower probability.
These findings are important because lifting the Sanitary Emergency Measures, firstly implemented in late March in Mexico City, is programmed for May 30, 2020.
If our scenarios are correct, the risk of a new outbreak is high, given that the date for ending confinement would coincide with the dates predicted to be of maximum incidence.
Moreover, even if maximum incidence occurs in early May 2020 and incidence decreases the following days, the number of susceptible individuals will still be large.
Since SARS-CoV-2 is a new virus, there is yet no significant herd immunity in the population.
In Mexico, April 30 (Children's day) and May 10 (Mother's day) are significant dates for family gatherings and celebrations.
To the date of submission, the effect of these perturbations on the epidemic curve are yet unknown.
However, our modelling approach allows for the consideration of these actions and the planning of mitigation or other intervention measures because of its probabilistic nature. 

Our model projects, namely, that peak incidence will likely occur in late May or early June 2020, together with the crucial lack of sufficient testing to provide a more accurate estimate of the number of people infected, provides support for recommending a reevaluation of the date, but also a gradual and slow release of mitigation and social-distancing measures to prevent a fast rebound of the epidemic.

As for future work, there are a number of avenues we are currently exploring, both theoretical from the modelling side and practical, as a predictive tool.
For instance, we will shortly explore the likely impact for Mexico City of lifting Sanitary Emergency measures too soon.
Clearly, we plan to extend this analysis to other regions of Mexico.

\acknow{RHM is grateful for the support of CONTEX project 2018-9B. JXVH acknowledges support from grant UNAM-DGAPA-PAPIIT IN115720. GACS kindly acknowledges support from UNAM-DGAPA-PAPIIT IN114717. LB acknowledges support from UNAM-DGAPA-PAPIIT IG100819. We also thank SECTEI-CDMX for providing data on the evolution of COVID-19 in Mexico City. We thank H\'ector Benitez, IIMAS-UNAM, for his unwavering support during the elaboration of this work. }

\showacknow{} 
\bibliography{pnas-sample}

\end{document}